\documentclass[aps,prl,twocolumn,superscriptaddress,showpacs,
preprintnumbers]{revtex4}
\usepackage{amsmath,amssymb}
\usepackage{graphicx}
\renewcommand\d{\partial}

\newcommand\x{\mathbf{x}}

\newcommand\+{\dagger}

\newcommand\boldnabla{\mbox{\boldmath$\nabla$}}

\preprint{INT-PUB-11-037}
\begin{document}
\title{Hall Viscosity and Electromagnetic Response}
\author{Carlos Hoyos}  
\affiliation{Department of Physics, University of Washington, Seattle, 
Washington 98195-1560, USA}
\author{Dam Thanh Son}
\affiliation{Institute for Nuclear Theory, University of Washington, Seattle,
Washington 98195-1550, USA}
\begin{abstract}
We show that, for Galilean invariant quantum Hall states,
 the Hall viscosity appears in the electromagnetic response 
at finite wave numbers $q$.  In particular, the leading $q$ dependence 
of the Hall conductivity 
at small $q$ receives a contribution from
the Hall viscosity.  The coefficient of the $q^2$ term in the
Hall conductivity is universal in the limit of strong magnetic field.
\end{abstract}
\pacs{73.43.Cd}
\maketitle

\emph{Introduction.}---Quantum Hall states have been shown to possess,
in addition to the Hall conductivity, a new property called the Hall
viscosity~\cite{ASZ:1995,Avron:1997}.  The Hall viscosity breaks
parity,  is dissipationless
and can be defined at zero temperature.  It has been
shown recently~\cite{Read:2008,ReadRezayi:2010} that the Hall
viscosity is related to a topological property of the quantum Hall
state---the Wen-Zee shift~\cite{WenZee:1992}.

One may ask how the Hall viscosity can be measured.  As originally
defined, the Hall viscosity is related to the stress response of the system to
metric perturbations.  Such perturbations can be, in principle,
mimicked by lattice vibrations (sound waves).  It has also been
suggested that the Hall viscosity determines the stress created by an
inhomogeneous electric field~\cite{Haldane:2009}.  In this paper we
show that, for quantum Hall states of systems with Galilean invariance
and made up of particles of the same charge/mass ratio, the Hall
viscosity can be, in principle, determined from electromagnetic
response alone.  We shall show this result first using intuitive
physical arguments, and then by employing the formalism of
nonrelativistic diffeomorphism invariance, applied to the low-energy
effective action of the Hall liquid.
\vspace{3pt}

\begin{figure}[ht]
\includegraphics[width=0.25\textwidth]{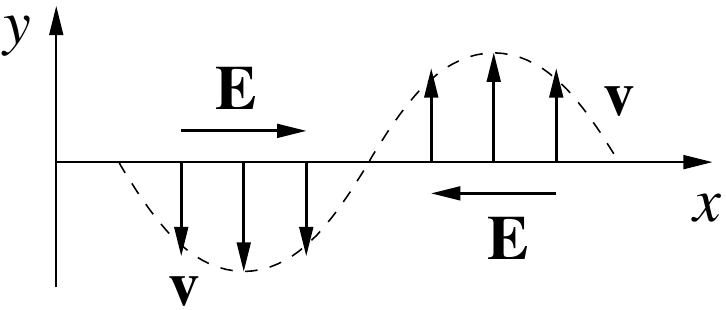}
\caption{Pattern of flow in an inhomogeneous electric field.}
\label{fig:flow}
\end{figure}

\emph{Main result.}---Consider a quantum Hall state in finite magnetic
field $B$.  First we concentrate on the case when the interaction
between particles is short-ranged.  (The case of Coulomb interaction
will be treated later in the paper.)  Let us turn on a static
longitudinal electric field ${\bf E}=-\boldnabla\phi$ where $\phi$ is
the scalar potential.  We take $\phi$ to vary in space with some wave
vector ${\bf q}$ pointing along the $x$ direction and measure
the Hall current $j_y$ (see Fig.~\ref{fig:flow}).  
The proportionality between $j_y$ and $E_x$
is the wave-vector dependent Hall conductivity,
\begin{equation}
  j_y(q) = \sigma_{xy}(q) E_x(q).
\end{equation}
In the limit $q\to0$, $\sigma_{xy}(q)$ approaches the universal value,
determined by the rational filling factor $\nu$: $\sigma_{xy}(0)=\nu
e^2/(2\pi\hbar)$.  In general, $\sigma_{xy}$ has a nontrivial
dependence on the wave number $q$.

We will show that, for a Galilean invariant system of electrons,
the coefficient $C_2$ of the first correction in
the low-$q$ expansion of the Hall conductivity 
\begin{equation}
  \frac{\sigma_{xy}(q)}{\sigma_{xy}(0)} = 1 + C_2 (q\ell)^2
  +{\cal O}(q^4\ell^4) , \label{C1}
\end{equation}
can be related to the Hall viscosity $\eta^a$ and the function $\epsilon(B)$
which is is the
energy density (energy per unit area) as function of the external
magnetic field $\epsilon(B)$ at fixed filling factor,
\begin{equation}
  C_2 =  \frac{\eta^a}{\hbar n} - \frac{2\pi}\nu \frac{\ell^2}{\hbar\omega_c}
     B^2 \epsilon''(B).  \label{C2}
\end{equation}
Here $\ell=\sqrt{\hbar c/|e|B}$ is the magnetic length,
$\omega_c=|e|B/mc$ is the cyclotron frequency, and $n$ is the density of
electrons.

Using the relationship between $\eta^a$
and the shift ${\cal S}$: $\eta^a=\hbar n {\cal
  S}/4$~\cite{Read:2008,ReadRezayi:2010}, the first term in the right hand side
of Eq.~(\ref{C2}) can be written as ${\cal S}/4$, which makes clear
that the magnitude of this contribution is robust (i.e., does not
depend on interactions).
The second contribution involves the function $\epsilon(B)$ and is not
universal.
However, its magnitude can be extracted independently by measuring 
currents created by weak inhomogeneous
perturbations of the magnetic field $\delta B$,
\begin{equation}
  {\bf j} = -c \epsilon''(B) \hat{\bf z} \times \boldnabla \delta B.
\end{equation}
Hence, by measuring the electromagnetic response of the system to
inhomogenous electric and magnetic fields, one can determine the Hall
viscosity.

The situation becomes simpler in the limit of high magnetic fields 
(i.e., that of no mixing between Landau levels) 
in which the energy $\epsilon(B)$ becomes that of
non-interacting electrons in a magnetic field.  For the integer quantum
Hall state with $\nu=N$, the energy density
$\epsilon(B)=(N^2/4\pi)\hbar\omega_c/\ell^2$, and the shift ${\cal
  S}=N$, so we have
\begin{equation}
  \frac{\sigma_{xy}(q)}{\sigma_{xy}(0)}
   =  1 - \frac{3N}4(q\ell)^2 + {\cal O}(q^4\ell^4)
   \quad \textrm{for $\nu=N$}.
\end{equation}
The result coincides with what has been computed in the literature
($\sigma_{xy}$ is proportional to $\Sigma_1$ in the notations of
Ref.~\cite{Chen:1989xs}).  For fractional quantum Hall states with $\nu<1$, $\epsilon(B)=(\nu/4\pi)\hbar\omega_c/\ell^2$, therefore $C_2=\frac14{\cal S}-1$.  In particular,
for Laughlin's states with
$\nu=1/(2k{+}1)$, the shift ${\cal S}=2k{+}1$~\cite{WenZee:1992}, so
\begin{equation}\label{FQH}
  \frac{\sigma_{xy}(q)}{\sigma_{xy}(0)} = 
   1+\frac{2k-3}4(q\ell)^2 + {\cal O}(q^4\ell^4),\quad \nu= \frac1{2k{+}1}
   \,.
\end{equation}
In general, for any quantum Hall state, we can find the $q^2$
correction to $\sigma_{xy}(q)$ from the value of the shift ${\cal S}$
and the total energy, as a function of the magnetic field.
\vspace{3pt}

\emph{Physical argument.}---Before presenting the mathematical proof
of the statement made above, we will give a physical derivation.  We
will show that the two contributions to $C_2$ come from two different
physical effects.

First let us note that to first approximation, the Hall fluid moves
along the $y$ direction with a velocity that depends on $x$ (see
Fig.~\ref{fig:flow}),
\begin{equation}\label{flow}
  v_y(x) = - \frac{c E_x(x)}B\,.
\end{equation}
This velocity is determined by balancing electric and magnetic forces
acting on a fluid volume.  However, the flow~(\ref{flow}) is a shear
flow with a nonzero strain rate.  The Hall viscosity leads to an additional
stress in the system, which in turn induces a correction to the current.

Let us compute the magnitude of the this effect. The strain rate
$V_{xy} = \frac12 \d_x v_y$ induces, through the Hall viscosity, 
an additional contribution to the stress, 
$\sigma_{xx} = -\sigma_{yy} = 2\eta^a V_{xy}$
The $x$-dependence of $\sigma_{xx}$ leads to an additional force
acting on each volume element of the fluid along the 
$x$ axis:
$f_x = - \d_x \sigma_{xx}$.
This force induces a correction to the Hall current equal to
\begin{equation}
  \delta j_y =  -\frac cB f_x = -\frac{\eta^a c^2}{B^2}E_x''(x) .
\end{equation}
We thus find the first correction to
$\sigma_{xy}$,
\begin{equation}
  \sigma_{xy}^{(1)}(q) = 
  \frac{\eta^a c^2}{B^2}q^2 .
\end{equation}

The second effect is related to the fact that the fluid flow, in
addition to having a shear rate, also has a nonzero local angular
velocity: 
\begin{equation}
  \Omega(x) = \frac12\d_x v_y = -\frac{cE_x'(x)}{2B}\,.
\end{equation}
This local rotation acts as an effective magnetic field, equal to
$\delta B = 2mc\Omega/e$ (found by equating the Coriolis force with the 
Lorentz force from $\delta B$.)
On the other hand, the quantum Hall fluid is a diamagnetic material.
with magnetic moment density $M=-\d\epsilon/\d B$.
For a constant magnetic field, $M$ is constant.  But due to the
fluctuations $\delta B$ there is an inhomogeneous contribution to
the magnetic moment density,
\begin{equation}
  \delta M = - \frac{\d^2\epsilon}{\d B^2} \delta B
          = \epsilon''(B)\frac{mc^2E_x'(x)}{eB}\,.
\end{equation}
This fluctuating magnetic moment density leads to an additional
electromagnetic current, ${\bf j}=c\,\hat {\bf z}\times\boldnabla M$:
\begin{equation}
  j_y = \epsilon''(B)\frac{mc^3E_x''(x)}{|e|B}\,.
\end{equation}
We find the second contribution to the Hall conductivity,
\begin{equation}
  \sigma_{xy}^{(2)}(q) = 
  - \frac{mc^3\epsilon''(B)}{|e|B}q^2.
\end{equation}
The finite-wave-number correction to the Hall conductivity is
$\sigma^{(1)}_{xy}+\sigma^{(2)}_{xy}$.  Elementary algebraic
manipulations bring it to the form of Eqs.~(\ref{C1}) and (\ref{C2}).
\vspace{3pt}

\emph{Diffeomorphism invariance.}---%
We now formally prove the result derived above by 
constructing a low-energy effective theory of the quantum Hall state.
As the quantum Hall state is gapped, the effective action is as
a local functional of the external fields.  Expanding in momentum to lowest order, it is simply
the Chern-Simons action. In order to reproduce the $q^2$ correction to
$\sigma_{\!xy}$ we need to go beyond leading order.

We shall make use of the nonrelativistic
diffeomorphism invariance, introduced in Ref.~\cite{Son:2005rv}.  Our
strategy is to couple our system to gravity and find out the symmetries
of the action.  These symmetries are inherited by the low-energy effective
theory, and impose nontrivial constraints
to the effective Lagrangian.

We consider a quantum Hall state in the presence of an external gauge field
$A_\mu(t,\x)$ and a spatial metric $g_{ij}(t,\x)$.  For example, for the case of
free fermions we assume the action to be
\begin{multline}\label{S-free}
  S_0 = \!\int\!{\rm d}t\,{\rm d}^2x\, \sqrt g \Bigl[ 
  \frac i2 (\psi^\+ \d_t \psi - \d_t \psi^\+ \psi) + A_0\psi^\+\psi
       \\   - \frac{g^{ij}}{2m} (\d_i \psi^\+ + i A_i \psi^\+)
        (\d_j \psi - i A_j \psi) \Bigr].
\end{multline}
We will set $\hbar=1$ and absorb an $e/c$ factor into the normalization of the gauge potential $A_i$.
Most of the time we will set the spatial metric to be flat
($g_{ij}=\delta_{ij}$) at the end of calculations, but it will be useful
to consider a general metric in the intermediate steps.

The action~(\ref{S-free}) is invariant under reparametrization of
spatial coordinates $x^k \to x^k+\xi^k$, where $\xi^k$ depends both on
space and time, $\xi^k=\xi^k(t,\x)$.  The passive form of the
transformations is
\begin{align}
  \delta A_0 &= -\xi^k \d_k A_0 - A_k \dot\xi^k, \label{GCT-A0} \\
  \delta A_i &= -\xi^k \d_k A_i - A_k \d_i \xi^k - m g_{ik}\dot \xi^k ,
    \label{GCT-Ai}\\
  \delta g_{ij} &= -\xi^k \d_k g_{ij} - g_{kj} \d_i \xi^k 
                   - g_{ik}\d_j \xi^k , \label{GCT-gij} \\
  \delta\psi &= -\xi^k \d_k \psi  \label{GCT-psi} .
\end{align}
The Galilean transformation is a special case with $\xi^k=v^k t$.  As
explained in Ref.~\cite{Son:2005rv}, the transformations above
can be motivated by taking a nonrelativistic limit of relativistic
diffeomorphisms.

Interactions can be introduced in a way which preserves the diffeomorphism
invariance.  For example, by adding to (\ref{S-free})
\begin{equation}
  S = S_0 + \!\int\!{\rm d}t\,{\rm d}^2x\, \sqrt g\, \Bigl(\lambda\psi^\+\psi\phi 
  - \frac12 g^{ij}\d_i\phi\d_j\phi - \frac{m_\phi^2}2\phi^2\Bigr)
\end{equation}
one introduces an attractive potential of range $m_\phi^{-1}$ between
the particles.  The new action is diffeomorphism invariant if $\phi$
transforms as a scalar $\delta\phi=-\xi^k\d_k\phi$.  A generic
potential decaying faster than an exponential can be represented by an
infinite number of mediating fields, and so coupling to the external
metric can be made compatible with diffeomorphism invariance.

Coulomb interactions can also be introduced, but now the field
mediating the interaction propagates in three spatial dimensions.  We can
assume that the spatial metric does not depend on the third direction
\begin{multline}
  S = S_0+ \int\!{\rm d}t\,{\rm d}^2x\,\sqrt{g}\, a_0(\psi^\+\psi-n_0)\\
  +\frac{2\pi\varepsilon}{e^2}\!\int\!{\rm d}t\,{\rm d}^2x\,{\rm d}z\, \sqrt g\,
   \bigl[g^{ij}\d_ia_0\d_ja_0 +(\d_za_0)^2\bigr].
\end{multline}
($\varepsilon$ is the dielectric constant).  We have included a
neutralizing background with density $n_0$.  The full action is
diffeomorphism invariant if $a_0$ transforms as a scalar: $\delta
a_0=-\xi^k\d_k a_0$.
\vspace{3pt}

\emph{Power counting.}---%
We now start constructing the
low-energy effective field theory of the quantum Hall states.  For
incompressible states, there is no low-energy excitations, and we can
integrate out $\psi$.  If interactions are short-ranged, the fields
$\phi$ mediating interactions can also be integrated out.  Thus the
effective Lagrangian is a local function of the external fields
$A_\mu$, $g_{ij}$ and their derivatives.  The effective action must be
invariant under (\ref{GCT-A0})---(\ref{GCT-gij}).

To organize a derivative expansion, one needs a power-counting scheme
with a small parameter.  There is an ambiguity in choosing the scheme,
as the time derivative $\d_t$ and spatial derivatives can be chosen to
be independent expansion parameters.  For definiteness, in this paper
we use the following scheme.  All quantities will be regarded as
proportional to some powers of a small parameter $\epsilon$, times
some powers of $\omega_c$ and $\ell$.  The external fields are assumed
to vary slowly in space and time,
\begin{equation}
  \d_i \sim \epsilon\ell^{-1}, \quad \d_t \sim \epsilon^2 \omega_c .
\end{equation}
As for the magnitude of external perturbations, we assume
\begin{equation}
  \delta A_0 \sim \epsilon^0 \omega_c, \quad
  \delta A_i \sim \epsilon^{-1} \ell^{-1},\quad
  \delta g_{ij} \sim 1 .
\end{equation}
In this scheme, we allow for order one variations of the metric, the
magnetic field ($\delta B\sim
\epsilon^0\ell^{-2}$) and the chemical potential ($A_0$).  In further
formulas, the electric and magnetic fields are defined as
\begin{equation}
  E_i=\d_i A_0-\d_0 A_i, ~
  B = \frac{F_{12}}{\sqrt g} = \frac{\epsilon^{ij}\d_i A_j}{\sqrt g} 
    \equiv \varepsilon^{ij}\d_i A_j,
\end{equation}
so $E_i=O(\epsilon)$ and $B=O(1)$.
\vspace{3pt}

\emph{Chern-Simons and Wen-Zee terms.}---%
Two important ingredients in our construction of the effective field
theory are the Chern-Simons action and the Wen-Zee action.  The Chern-Simons
action is
\begin{equation}
  S_{\rm CS} = \frac\nu{4\pi}\!\int\!{\rm d}t\,{\rm d}^2x\, 
       \epsilon^{\mu\nu\lambda} A_\mu\d_\nu A_\lambda\,,
\end{equation}
and is of order $\epsilon^0$ in our power counting scheme.  This will
be the leading term in the effective action.  To construct the Wen-Zee
action, we first define the spin connection.  We introduce a spatial
vielbein $e^a_i$, $a=1,2$ so that $g_{ij} = e^a_i e^a_j$ and
$\epsilon^{ab}e^a_i e^b_j=\varepsilon_{ij}$.  The vielbein is defined
up to local O(2) rotations in $a$ space.  If we now define the
connection $\omega_\mu$,
\begin{align}
  \omega_0 &= \frac12\epsilon^{ab} e^{aj}\d_0 e^b_j, \\
  \omega_i &= \frac12\epsilon^{ab} e^{aj}\nabla_{\!i} e^b_j
           = \frac12 (\epsilon^{ab} e^{aj}\d_i e^b_j
               -\varepsilon^{jk}\d_j g_{ik}),
\end{align}
then under local O(2) rotations $\omega_\mu$ transforms like an
Abelian gauge potential $\omega_\mu \to \omega_\mu -\d_\mu \lambda$.
By using $\omega_\mu$ we can construct the following gauge invariant
action
\begin{equation}
  S_{\rm WZ} = \frac\kappa{2\pi}\!\int\!{\rm d}t\, {\rm d}^2x\,
     \epsilon^{\mu\nu\lambda}\omega_\mu \d_\nu A_\lambda .
\end{equation}
This action is of order $\epsilon^2$ in our power counting scheme and
has to be included if we work to that order.  The $\omega d\omega$
Chern-Simons term, on the other hand, is of order $\epsilon^4$ and
will not be considered.

The coefficient $\kappa$ is related to the shift ${\cal S}$.  Indeed,
the ``torsion magnetic'' field
$\d_1\omega_2-\d_2\omega_1=\frac12\sqrt g R$ where $R$ is the
scalar curvature.  Integrating by parts, the Wen-Zee action contains a
term
\begin{equation}
  \frac{\kappa}{2\pi}\epsilon^{\mu\nu\lambda} \omega_\mu \d_\nu A_\lambda 
  \simeq \frac\kappa{4\pi}\sqrt g\, A_0 R + \cdots
\end{equation}
which gives a contribution to the particle number density that is
proportional to the scalar curvature.  If the quantum Hall state lives on
a closed two dimensional surface, then the total number of particles is
\begin{equation}\label{QNchi}
  Q= \!\int\!{\rm d}^2x\, \sqrt g\, j^0 
   = \!\int\!{\rm d}^2x\, \sqrt g\left( \frac\nu{2\pi} B 
     {+} \frac\kappa{4\pi} R\right)
   = \nu N_\phi {+} \kappa \chi
\end{equation}
where $N_\phi$ is the total number of magnetic fluxes and
$\chi=2(1-g)$ is the Euler character.  Comparing to the definition of
${\cal S}$ in Ref.~\cite{WenZee:1992}, we find
$\kappa=\frac12\nu{\cal S}$.
For the integer Quantum Hall state with $\nu=N$, $\kappa=N^2/2$.  For
Laughlin's states $\kappa=1/2$.

The Wen-Zee action gives rise to Hall viscosity~\cite{GoldbergerRead}.
Expanding the WZ term to quadratic order, one finds, among other
terms,
\begin{equation}
  S_{\rm WZ} = -\frac{\kappa B}{16\pi}\epsilon^{ij} \delta g_{ik}\d_t \delta g_{jk}
  + \cdots
\end{equation}
which implies the presence of an odd term in the stress tensor two
point function, or Hall viscosity.  The value of the Hall viscosity is
$\eta^a = \kappa B/4\pi= \frac14 {\cal S} n$.  This relationship
between the Hall viscosity and the shift has been derived previously
in Ref.~\cite{ReadRezayi:2010}.
\vspace{3pt}

\emph{Most general effective action.}---%
It is straightforward to verify that both $S_{\rm CS}$ and $S_{\rm
  WZ}$ are not diffeomorphism invariant, and need to be corrected.  In
fact, to order $O(\epsilon^2)$, the most general effective action can
be written as $S=\int\!{\rm d}t\,{\rm d}^2x\,\sqrt g\, \sum_{i=1}^5
{\cal L}_i$, where ${\cal L}_i$ ($i=1,\ldots,5$) are five independent general
diffeomorphism invariant (to order $\epsilon^2$) terms
\begin{align}
  {\cal L}_1 &= \frac\nu{4\pi}\! 
         \Bigl( \varepsilon^{\mu\nu\lambda} A_\mu \d_\nu A_\lambda
      +  \frac mB g^{ij} E_i E_j\Bigr),\\
  {\cal L}_2 & = \frac{\kappa}{2\pi}\! 
          \Bigl(\varepsilon^{\mu\nu\lambda}\omega_\mu\d_\nu A_\lambda
          + \frac1{2B}\, g^{ij} \d_i B\, E_j\Bigr),\\ 
  {\cal L}_3 &= -
     \epsilon(B) - \frac mB \epsilon''(B) g^{ij}\d_i B\, E_j,\\
  {\cal L}_4 &= -\frac12 
          K(B) g^{ij}\d_i B\,\d_j B, \\
  {\cal L}_5 &=  R\, h(B),
\end{align}
where $\epsilon(B)$, $K(B)$, and $h(B)$ are functions of $B$.  The
function $\epsilon(B)$ has the physical meaning of the energy density
of the quantum Hall state as a function of the magnetic field $B$,
${\cal L}_4$ and ${\cal L}_5$ do not enter the quantities of of our
interest.  The next to leading order term in ${\cal L}_1$ enforces
compliance with Kohn's theorem.  The two-point function of the
electromagnetic current $j^\mu$ is obtained by taking the second
derivative of the effective action with respect to $A_\mu$, then
setting perturbations to zero.  Equivalently we can differentiate the
effective action once with respect to the external fields to get the
current. We find, in flat space
\begin{equation}
  j^i = \frac\nu{2\pi}\epsilon^{ij}E_j - \frac1B\left[\frac\kappa{4\pi}
    - m\epsilon''(B)\right]\epsilon^{ij}\d_j(\boldnabla\!\cdot\!{\bf E})
    + \cdots 
\end{equation}
where $\cdots$ refers to term that vanish when the magnetic field is
not perturbed.  Equations~(\ref{C1}) and (\ref{C2}) are reproduced from
this formula.
\vspace{3pt}

\emph{Inclusion of Coulomb interactions}.---In the case with Coulomb 
interactions, 
one needs to take into account the screening of the electric field.
The expansion~(\ref{C1},\ref{C2}) therefore applies not to 
$\sigma_{xy}(q)$ but to
\begin{equation}
  \tilde\sigma_{xy}(q) = \left[ 1+ \frac{e^2\chi(q)}{2\pi\epsilon\, q}\right]
   \sigma_{xy}(q)
  \simeq \left[1+\frac{\nu\varkappa}\pi (q\ell)\right]
    \sigma_{xy}(q)
\end{equation}
where $\varkappa=e^2/(4\pi\epsilon\ell\omega_c)$ and $\chi(q)$ is the
static susceptibility, the small-$q$ behavior of which is is
determined by Kohn's theorem: $\chi(q)=\nu m q^2/(2\pi B)$.  In the
limit of high magnetic fields where $\varkappa\ll1$, the distinction
between $\sigma_{xy}$ and $\tilde\sigma_{xy}$ disappears.
\vspace{3pt}

\emph{Conclusions.}---%
We have shown that the Hall viscosity does not only
appear in the response to gravitational fluctuations, but also, under certain
circumstances, in a purely electromagnetic response function.  For this
one needs Galilean invariance and that all particles have the same
charge/mass ratio, a condition satisfied in the most interesting
physical systems.

One notes that topological arguments alone are insufficient to
determine the coefficient of the $q^2$ term in the finite wave number
Hall conductivity.  But topology, coupled with nonrelativistic
diffeomorphism invariance, is powerful enough to find this coefficient
[e.g., Eq.~(\ref{FQH})].  It would be interesting to explore
consequences of diffeomorphism invariance for other systems with
topological order, e.g., the $p_x+ip_y$ paired state or the superfluid
B phase of $^3$He or the compressible $\nu=1/2$ state.

Finally, the wave number dependence of the Hall conductivity should be
measured and checked against our prediction.  Such a measurement would
be a measurement of the Hall viscosity.

We thank T.~Hughes, X.~Qi, E.~Witten for stimulating discussions, and
N.~Read for explaining the connection between the Wen-Zee term and
Hall viscosity.  This work is supported, in part, by DOE grants
DE-FG02-96ER40956 and DE-FG02-00ER41132 and NSF grant No.\ NSF PHY05-51164.

After this work was finished, we learned that the first contribution
on the right hand side of Eq.~(\ref{C2}) has been derived by
B.~Bradlyn, M.~Goldstein, and N.~Read~\cite{BGR}. We thank N.~Read for
communicating this result to us.

\end{document}